\documentclass[aps,pre,reprint,superscriptaddress]{revtex4-1}
\usepackage[latin9]{inputenc}
\usepackage{prettyref}
\usepackage{float}
\usepackage{amsmath}
\usepackage{amsthm}
\usepackage{amssymb}
\usepackage{graphicx}
\begin{document}
\title{Temperature expressions and ergodicity\\ of the Nos\'e-Hoover deterministic
schemes}
\author{A. Samoletov}
\email{A.Samoletov@liverpool.ac.uk}
\affiliation{Department of Mathematical Sciences, The University of Liverpool,
UK}
\affiliation{Institute for Physics and Technology, Ukraine}
\author{B. Vasiev}
\email{B.Vasiev@liverpool.ac.uk}
\affiliation{Department of Mathematical Sciences, The University of Liverpool,
UK}
\begin{abstract}
Thermostats are dynamic equations used to model thermodynamic variables in molecular dynamics. The applicability of thermostats is based on the ergodic hypothesis. The most commonly used thermostats are designed according to the Nos\'e-Hoover scheme, although it is known that it often violates ergodicity. Here, following a method from our recent study \citep{SamoletovVasiev2017}, we have extended the classic Nos\'e-Hoover scheme with an additional temperature control tool. However, as with the NH scheme, a single thermostat variable is used. In the present study we analyze the statistical properties of the modified equations of motion with an emphasis on ergodicity. Simultaneous thermostatting of all phase variables with minimal extra computational costs is an advantage of the specific theoretical scheme presented here.
\end{abstract}
\maketitle

\section{Introduction\label{sec:Introduction}}

Molecular dynamics (MD)  \citep{AllenTildesley1989,frenkel2002understanding,tuckerman2010statistical,LeimkuhlerMatthews2015,hoover2012computational} is an essential part of research in a range of disciplines in natural sciences and in engineering including such popular branches as the design of new functional materials and the drug discovery. MD simulations are performed under certain thermodynamic conditions, typically at fixed temperature or pressure. Many different dynamic temperature control tools (thermostats), deterministic and stochastic, have been proposed  \citep{LeimkuhlerMatthews2015,hoover2012computational,SamoletovDettmannChaplain2007,SamoletovDettmannChaplain2010,LeimkuhlerNoorizadehTheil2009,JeppsRondoni2010}. Recently, we have shown that a range of thermostats can be derived in the framework of a unified approach based on the fundamental principles of statistical physics \citep{SamoletovVasiev2017}. However, this result has been presented in a rather formal theoretical form, so the benefits of the unified approach presented may not seem obvious in terms of practical use. To address this, particular cases of abstract results can be compared with well-known thermostat schemes though obviously with loss of mathematical generality.

The Nos\'e--Hoover (NH) method \citep{nose1984molecular,hoover1985canonical}) is commonly used in applications. This deterministic thermostat allows the canonical distribution for modeled physical system to be obtained by means of a single extra degree of freedom. The usability of NH schemes relies on the ergodic hypothesis which claims that a physical phase space trajectory will spend an equal amount of time in each phase space volume of equal probability  \citep{khinchin1949mathematical,arnol1968ergodic}. In other words, this hypothesis equates the long-time average of a physical observable to its ensemble average. While it is known that deterministic NH thermostats often violate ergodicity, they are assumed to be applicable for practical purposes. In order to improve the ergodicity of the NH method, a series of modifications have been proposed, \emph{e.g.} \citep{hamilton1990modified,kusnezov1990canonical,martyna1992nose,HooverHolian1996,PatraBhattacharya2014,Sprott2018}. At the same time, many studies have been done on the ergodicity violation of NH thermostats applied to low-dimensional systems, primarily to the harmonic oscillator  \citep{Hoover2016,HooverSprottPatra2015,artemov2019cumulant,PatraBhattacharya2014a,watanabe2007ergodicity}, where rigorous results have been obtained \citep{legoll2007non,legoll2009non}. 

Characteristic features of the NH thermostat are 1) it is deterministic; 2) a single extra variable is added; 3) kinetic temperature is used. In the NH scheme, the extra variable is introduced in a formal way, so that its physical interpretation is missing. In contrast, the theoretical scheme from \citep{SamoletovVasiev2017} is firmly based on the fundamental laws of statistical physics. It is assumed that the physical system, $\mathrm{S}$, placed in the thermal reservoir, $\Sigma$, should to some extent perturb it and will itself be affected by the backward influence of this perturbation. Thus, the thermal reservoir is naturally divided into two parts, namely, the part that is involved in joint dynamics with system, $\mathrm{S}^{*}$, and the unperturbed part, $\Sigma\setminus\mathrm{S}^{*}$, staying in permanent thermal equilibrium. An important assumption made is that all the systems involved in a joint dynamics are statistically independent at equilibrium. In such a framework, the extra variable is related to the perturbed part, $\mathrm{S}^{*}$, of the thermal reservoir. Therefore, a dynamic temperature control related to this extra degree of freedom is as fundamental as the kinetic energy of $\mathrm{S}$ system.

All of the above poses the question as to whether an NH type dynamics, which includes temperature expressions related to both systems, $\mathrm{S}$ and $\mathrm{S}^{*}$ and involves a single additional dynamic variable, is able to improve the ergodicity of the thermostat. We show in the study presented that the answer to this question is positive.

\section{Temperature expressions and thermostats}

As we have shown in \citep{SamoletovVasiev2017} thermostat schemes are based on the notion of temperature expressions. Let the system at hand, $\mathrm{S}$, when it is isolated, be defined by the phase space $\mathcal{M}$, the Hamiltonian function $H(x),\;x\in\mathcal{M}$, and equations of motion $\dot{x}=\boldsymbol{J}_{x}\mathbb{\boldsymbol{\nabla}}_{x}H(x)$, and similarly, let correspondingly $\mathrm{S}^{*}$ be defined by $\mathcal{M}^{*}$, $H^{*}(y),\;y\in\mathcal{M}^{*}$, and $\dot{y}=\boldsymbol{J}_{y}\boldsymbol{\nabla}_{y}H^{*}(y)$. Here, $\boldsymbol{J}_{x}$ and $\boldsymbol{J}_{y}$ are symplectic units. The function
of  system state, $\Theta(x,\vartheta),\:x\in\mathcal{M}$, is called a temperature expression if it explicitly depends on the temperature ($\vartheta=k_{B}T$, where $T$ is the temperature and $k_{B}$ is the Boltzmann constant) and satisfies the condition, 
\[
\int_{\mathcal{M}}\Theta(x,\vartheta)d\mu_{\vartheta}(x)=0\quad\textrm{for all}\quad \vartheta>0,
\]
$d\mu_{\vartheta}(x)=\rho_{\vartheta}(x)dx$, where $\rho_{\vartheta}(x)\propto exp\{-\vartheta^{-1}H(x)\}$ is the canonical (Gibbs) distribution. Similarly, 
\[
\int_{\mathcal{M}^{*}}\Theta^{*}(y,\vartheta)d\mu_{\vartheta}^{*}(y)=0\quad\textrm{for all}\quad \vartheta>0,
\]
where $d\mu_{\vartheta}^{*}(y)\propto exp\{-\vartheta^{-1}H^{*}(y)\}dy=\rho_{\vartheta}^{*}(y)dy$.
The ergodic hypothesis implies that for invariant densities $\rho_{\vartheta}(x)$
and $\rho_{\vartheta}^{*}(y)$,
\begin{alignat*}{1}
\lim_{T\rightarrow\infty}\frac{1}{T}\int_{0}^{T}\begin{cases}
\Theta(x(t),\vartheta)dt & =0\\
\Theta^{*}(y(t),\vartheta)dt & =0
\end{cases}
\end{alignat*}
for almost all trajectories. 

Among others, we have arrived at deterministic equations of motion
involving temperature expressions related to the $\mathrm{S}$ system 
($x$-variables) as well as to the
$\mathrm{S}^{*}$ system ($y$-variables) (\citep{SamoletovVasiev2017},
Section 3.B),
\begin{gather}
\dot{x}=\boldsymbol{J}_{x}\mathbb{\boldsymbol{\nabla}}_{x}H(x)+\sum_{\left(k\right)}\Theta_{k}^{*}(y,\vartheta)\boldsymbol{\varphi}_{k}(x),\nonumber \\
\dot{y}=\boldsymbol{J}_{y}\boldsymbol{\nabla}_{y}H^{*}(y)-\sum_{\left(l\right)}\Theta_{l}(x,\vartheta)\boldsymbol{\varphi}_{l}^{*}(y),\label{eq:GenNH}
\end{gather}
where $\left\{ \boldsymbol{\varphi}_{k}(x)\right\} _{(k)}$ is a set of vector fields on $\mathcal{M}$, $\left\{ \boldsymbol{\varphi}_{l}^{*}(y)\right\} _{(l)}$ is a set of vector fields on $\mathcal{M}^{*}$, and the following temperature expressions are chosen,
\begin{align}
\Theta_{l}(x,\vartheta)= & \boldsymbol{\boldsymbol{\varphi}}_{l}(x)\cdot\boldsymbol{\nabla}_{x}H(x)-\vartheta\boldsymbol{\nabla}_{x}\cdot\boldsymbol{\varphi}_{l}(x),\\
\Theta_{k}^{*}(y,\vartheta)= & \boldsymbol{\boldsymbol{\varphi}}_{k}^{*}(y)\cdot\boldsymbol{\nabla}_{y}H^{*}(y)-\vartheta\boldsymbol{\nabla}_{y}\cdot\boldsymbol{\varphi}_{k}^{*}(y).
\end{align}
The canonical density
\begin{gather}
\rho_{\infty}\propto\rho_{\vartheta}(x)\rho_{\vartheta}^{*}(y)=exp\{-\vartheta^{-1}[H(x)+H^{*}(y)]\}\label{eq:Canon-density}
\end{gather}
is invariant for dynamics \eqref{eq:GenNH}, provided that $\boldsymbol{\boldsymbol{\varphi}}_{k}^{*}(y)\exp[-\vartheta^{-1}H^{*}(y)]\underset{}{\rightarrow}\boldsymbol{0}$ as $\left|y\right|\rightarrow\infty$ and $\boldsymbol{\varphi}_{l}(x)\exp[-\vartheta^{-1}H(x)]\rightarrow\boldsymbol{0}$ as $\left|x\right|\rightarrow\infty$. The Liouville equation associated with the system \eqref{eq:GenNH} has the explicit form $\partial_{t}\rho=-\mathcal{L}^{*}\rho$, where $\mathcal{L}^{*}\rho=\boldsymbol{\nabla}_{x}\cdot(\dot{x}\rho)+\boldsymbol{\nabla}_{y}\cdot(\dot{y}\rho)$. Invariant probability densities are determined by the equation $\mathcal{L}^{*}\rho=0$. The proof that \emph{$\mathcal{L}^{*}\rho_{\infty}=0$} is by direct calculation \citep{SamoletovVasiev2017}.

The classical NH thermostat scheme is a particular case of dynamical system \eqref{eq:GenNH} corresponding to the specific selection of vector fields $\boldsymbol{\varphi}$ , $\boldsymbol{\varphi}^{*}$, and the Hamiltonian function $H^{*}$. For example, for a physical system with one degree of freedom, $x=\left(p,q\right)\in\mathbb{R}^{2}$,
\begin{equation}
H\left(p,q\right)=\frac{1}{2m}p^{2}+V\left(q\right),\label{eq:HanS}
\end{equation}
the system $\mathrm{S}^{*}$, $y=(\lambda,\xi)\in\mathbb{R}^{2}$,
\begin{gather}
H^{*}=\frac{1}{2Q}\lambda^{2},\label{eq:HamS2}
\end{gather}
and $\boldsymbol{\varphi}^{*}(y)=\left(-Q,0\right)$,  $\boldsymbol{\varphi}(x)=\left(p,0\right)$, where $Q>0$ is
a parameter, we arrive at the classical NH equations,
\begin{align}
\dot{q}= & \frac{p}{m},\nonumber \\
\dot{p}= & -V'(q)-\lambda p,\label{eq:NH classic}\\
\dot{\lambda}= & Q\left(\frac{p^{2}}{m}-\vartheta\right).\nonumber 
\end{align}
The invariant density corresponding to system \eqref{eq:NH classic} has the form of \eqref{eq:Canon-density}. The proof is by direct calculation. However, the dynamics \eqref{eq:NH classic} appear to be non-ergodic for this density \citep{Hoover2016,HooverSprottPatra2015,artemov2019cumulant,PatraBhattacharya2014a,watanabe2007ergodicity,legoll2007non,legoll2009non}. We can reformulate NH dynamics in a slightly generalized form with $H^{*}=h(\lambda)$, where $h(\lambda)$ is a suitable function and then arrive at the equations of motion 
\begin{alignat}{1}
\dot{q}= & \frac{p}{m},\nonumber \\
\dot{p}= & -V'(q)-Qh'(\lambda)p,\label{eq:NH classic2}\\
\dot{\lambda}= & Q\left(\frac{p^{2}}{m}-\vartheta\right).\nonumber 
\end{alignat}
The probability density \eqref{eq:Canon-density} is invariant for this dynamics \citep{fukuda2002tsallis,FukudaMoritsugu2015,bravetti2016thermostat}, while the ergodicity property is still questionable \citep{tapias2017ergodicity,Artemov2019FTVD}.
Note that thermostats \eqref{eq:NH classic} and \eqref{eq:NH classic2} include only the kinetic temperature expression and a single extra variable.

\section{Extended temperature control}

To answer the question posed at the end of \prettyref{sec:Introduction}, we consider the dynamical equations \eqref{eq:GenNH} under the following sets of vector fields $\boldsymbol{\varphi}$ and $\boldsymbol{\varphi}^{*}$ , supposing that both $\mathrm{S}$ and $\mathrm{S}^{*}$ are systems with one degree of freedom,
\begin{alignat*}{2}
\boldsymbol{\varphi}_{1}(x) & =\left(p,0\right), & \boldsymbol{\varphi}_{1}^{*}(y) & =\left(-Q,0\right);\\
\boldsymbol{\varphi}_{2}(x) & =\left(\varphi_{p}(q),0\right), & \quad\boldsymbol{\varphi}_{2}^{*}(y) & =\left(\lambda,0\right);\\
\boldsymbol{\varphi}_{3}(x) & =\left(0,\varphi_{q}(p)\right), & \quad\boldsymbol{\varphi}_{3}^{*}(y) & =\left(\lambda,0\right);
\end{alignat*}
where $\varphi_{q}(p)$ and $\varphi_{p}(q)$ are arbitrary functions that can be set by our choice. With Hamiltonian functions \eqref{eq:HanS} and \eqref{eq:HamS2} we arrive at the equations of motion,
\begin{align}
\dot{q}= & \frac{p}{m}+\varphi_{q}(p)\left(\frac{\lambda^{2}}{Q}-\vartheta\right),\nonumber \\
\dot{p}= & -\frac{dV(q)}{dq}-\lambda p+\varphi_{p}(q)\left(\frac{\lambda^{2}}{Q}-\vartheta\right),\label{eq:NH2}\\
\dot{\lambda}= & -\varphi_{q}(p)\frac{dV(q)}{dq}\lambda-\varphi_{p}(q)\frac{p}{m}\lambda+Q\left(\frac{p^{2}}{m}-\vartheta\right).\nonumber 
\end{align}
The required conditions are all fulfilled for this dynamics, that is, dynamical equations \eqref{eq:NH2} involve: 1) a single thermostat dynamic variable; 2) the kinetic temperature expression related to $\mathrm{S}$ system; 3) the density \eqref{eq:Canon-density} is invariant for dynamics \eqref{eq:NH2} (this is easily verified by direct calculation); and 4) the classical NH dynamics is a particular case of equations \eqref{eq:NH2} (indeed, with $\varphi_{q}(p)\equiv0$ and $\varphi_{p}(q)\equiv0$ we obtain system \eqref{eq:NH classic}). 

Furthermore, when $H^{*}=h(\lambda)$ we straightforwardly obtain the following equations,
\begin{align}
\dot{q}= & \frac{p}{m}+\varphi_{q}(p)\left(\lambda\frac{dh(\lambda)}{d\lambda}-\vartheta\right),\nonumber \\
\dot{p}= & -\frac{dV(q)}{dq}-\frac{dh(\lambda)}{d\lambda}p+\varphi_{p}(q)\left(\lambda\frac{dh(\lambda)}{d\lambda}-\vartheta\right),\label{eq:NH2h}\\
\dot{\lambda}= & -\varphi_{q}(p)\frac{dV(q)}{dq}\lambda-\varphi_{p}(q)\frac{p}{m}\lambda+\left(\frac{p^{2}}{m}-\vartheta\right),\nonumber 
\end{align}
so that the probability density \eqref{eq:Canon-density} is invariant for dynamics \eqref{eq:NH2h}.  $\Theta_{\lambda}(\lambda,\vartheta)=\lambda h'(\lambda)-\vartheta$ is a temperature expression related to the system $\mathrm{S}^{*}$. 

Note that the following specific forms of the $\varphi$-functions,
\begin{gather}
\varphi_{p}(q)\propto\left(q\frac{dV(q)}{dq}-\vartheta\right),\quad\varphi_{q}(p)\propto\left(\frac{p^{2}}{m}-\vartheta\right),\label{eq:DoubleTE}
\end{gather}
include virial \citep{hamilton1990modified} and kinetic temperature expressions.

\section{Numerical experiments}
\begin{figure}[H]
\includegraphics[width=8.5cm]{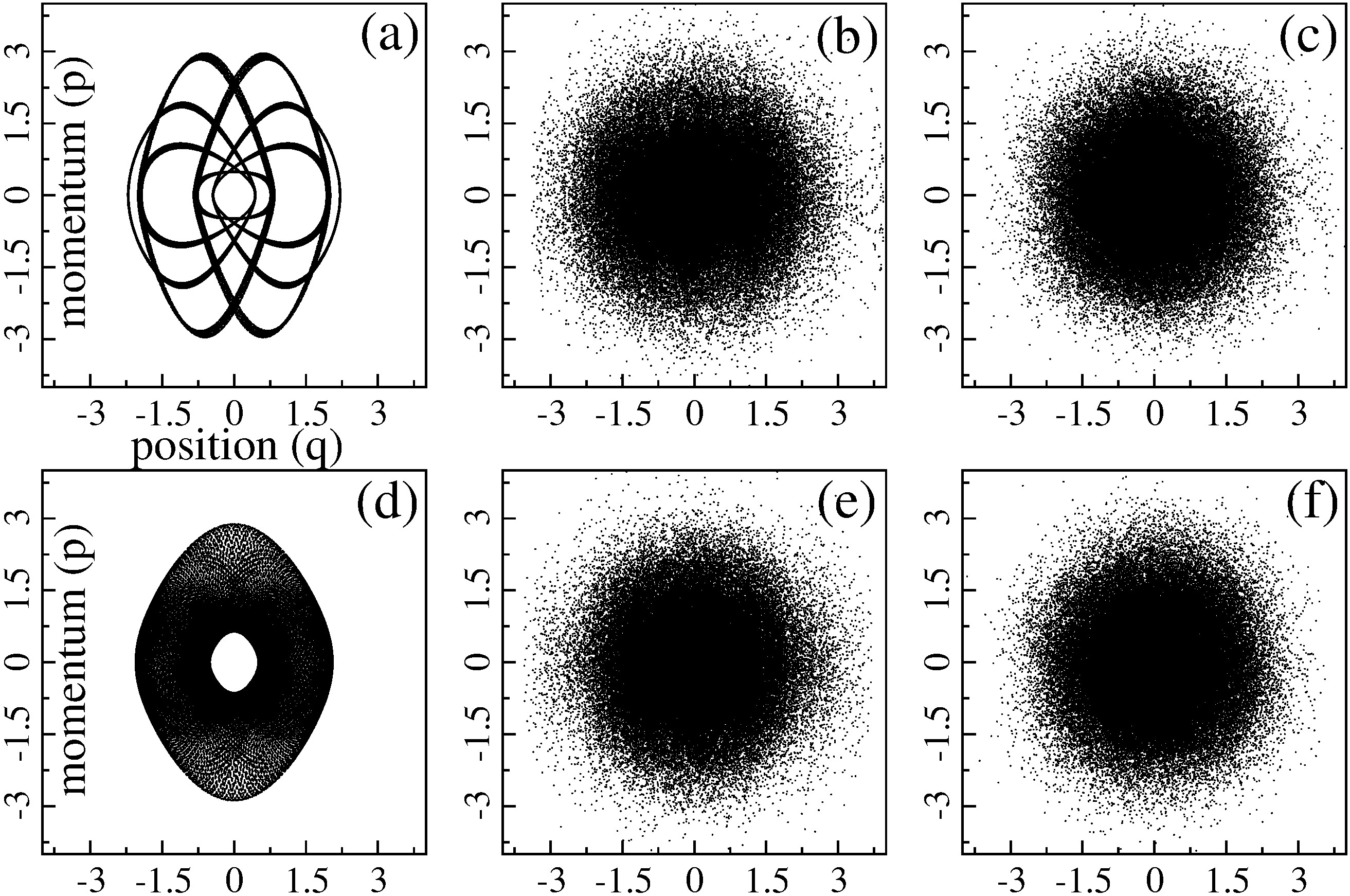}
\caption{\label{Fig.1} Computed trajectories (up to $t=10^{4}$) on the $(p,q)$ plane of a harmonic oscillator. Sub-pictures are arranged according to the nomenclature of numeric experiments as described above, that is,  {\bf{(a)}}=\{$\boldsymbol{\boldsymbol{\varphi}1},\mathbf{i1}$\}, {\bf{(b)}}=\{$\boldsymbol{\boldsymbol{\varphi}2},\mathbf{i1}$\}, {\bf{(c)}}=\{$\boldsymbol{\boldsymbol{\varphi}3},\mathbf{i1}$\}, 
	{\bf{(d)}}=\{$\boldsymbol{\boldsymbol{\varphi}1},\mathbf{i2}$\},
	{\bf{(e)}}=\{$\boldsymbol{\boldsymbol{\varphi}2},\mathbf{i2}$\},
	{\bf{(f)}}=\{$\boldsymbol{\boldsymbol{\varphi}3},\mathbf{i2}$\}.}
\end{figure}

\begin{figure}[H]
\includegraphics[width=8.5cm]{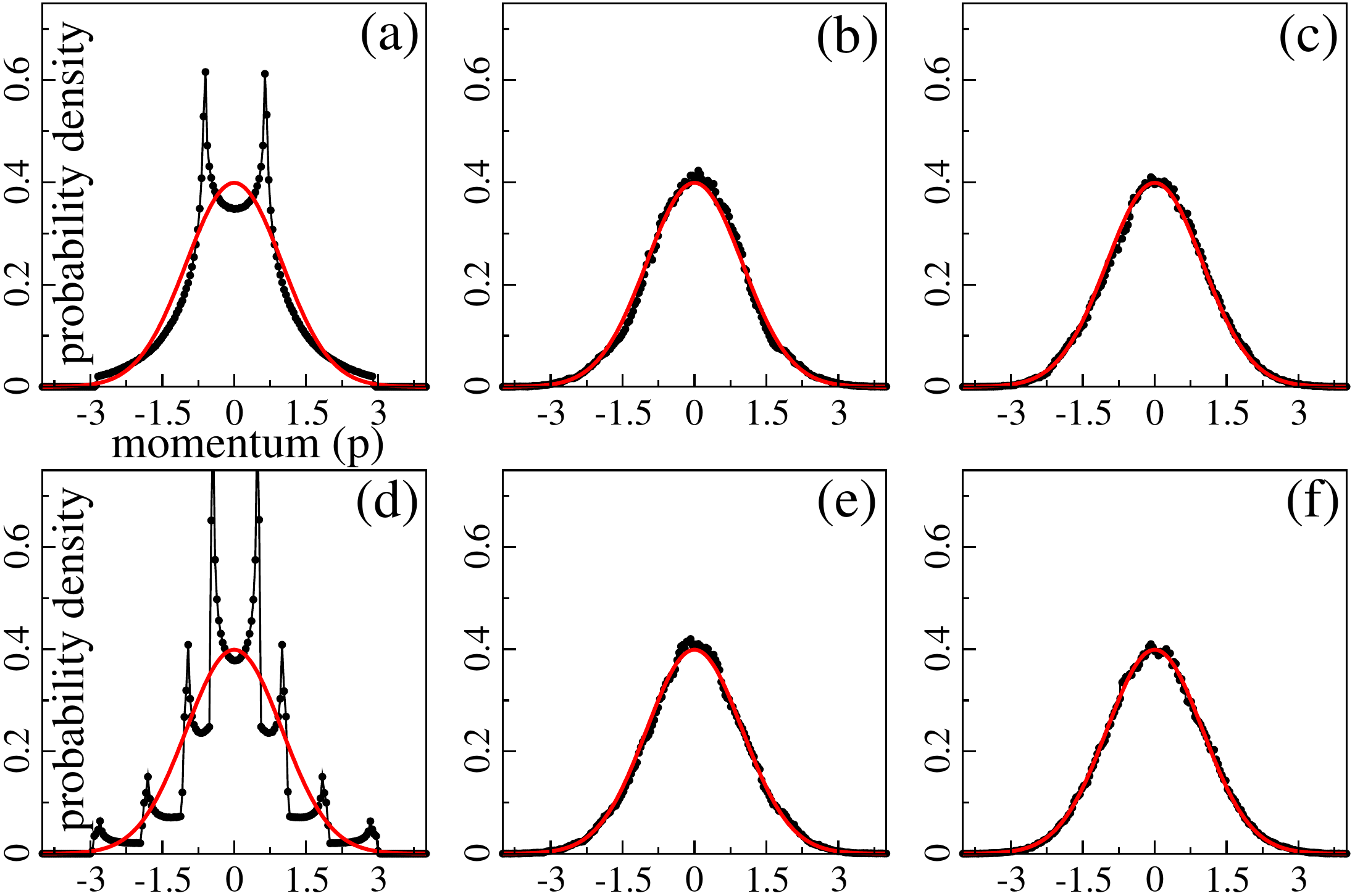}
\caption{\label{Fig.2} (Color online) The probability densities of the momentum variable $ p $ computed for the trajectories shown in Fig.~\ref{Fig.1}. The corresponding theoretical densities are shown in red.}
\end{figure}

In this section, we numerically compare the two forms of NH thermostat equations: the classic NH equations \eqref{eq:NH classic} and the advanced temperature control equations \eqref{eq:NH2}. As a model physical system we take the harmonic oscillator, $V\left(q\right)=\frac{1}{2}kq^{2},$ the system that has a well-documented problem of ergodicity. In what follows we explicitly compare three sets of functions $\left\{ \varphi_{p}(q),\varphi_{q}(p)\right\} $: 
\begin{description}
\item [{$\boldsymbol{\varphi}$1}] $\left\{ \varphi_{p}(q)\equiv0,\varphi_{q}(p)\equiv0\right\} $
(classical NH thermostat);
\item [{$\boldsymbol{\varphi}$2}] $\left\{ \varphi_{p}(q)=q,\varphi_{q}(p)=p\right\} $;
\item [{$\boldsymbol{\varphi}$3}] $\left\{ \varphi_{p}(q)=\left(qV'(q)-\vartheta\right),\varphi_{q}(p)=\left(\frac{1}{m}p^{2}-\vartheta\right)\right\} $; 
\end{description}
for two sets of initial conditions,
\begin{description}
\item [{i1}] q(0)=0.5, p(0)=0, $\lambda$(0)=0;
\item [{i2}] q(0)=0, p(0)=0.5, $\lambda$(0)=0;
\end{description}
\emph{ceteris paribus}, which we set equal to unity, that is, $m=1$, $k=1$, $Q=1$, $\vartheta=1$. We generate phase-space trajectories of length $t=10^{4}$ using the fourth-order Runge-Kutta method with time step $\Delta t=0.01$. 
We have performed a series of six numerical experiments corresponding to the following combinations of $\boldsymbol{\varphi}$-functions and initial conditions: 
{\bf{(a)}}=\{$\boldsymbol{\boldsymbol{\varphi}1},\mathbf{i1}$\}; 
{\bf{(b)}}=\{$\boldsymbol{\boldsymbol{\varphi}2},\mathbf{i1}$\}; 
{\bf{(c)}}=\{$\boldsymbol{\boldsymbol{\varphi}3},\mathbf{i1}$\}; 
{\bf{(d)}}=\{$\boldsymbol{\boldsymbol{\varphi}1},\mathbf{i2}$\}; 
{\bf{(e)}}=\{$\boldsymbol{\boldsymbol{\varphi}2},\mathbf{i2}$\}; 
{\bf{(f)}}=\{$\boldsymbol{\boldsymbol{\varphi}3},\mathbf{i2}$\}. 
The results of these experiments are shown in Fig.~\ref{Fig.1} and Fig.~\ref{Fig.2}.

\begin{figure}[H]
\includegraphics[width=8.5cm]{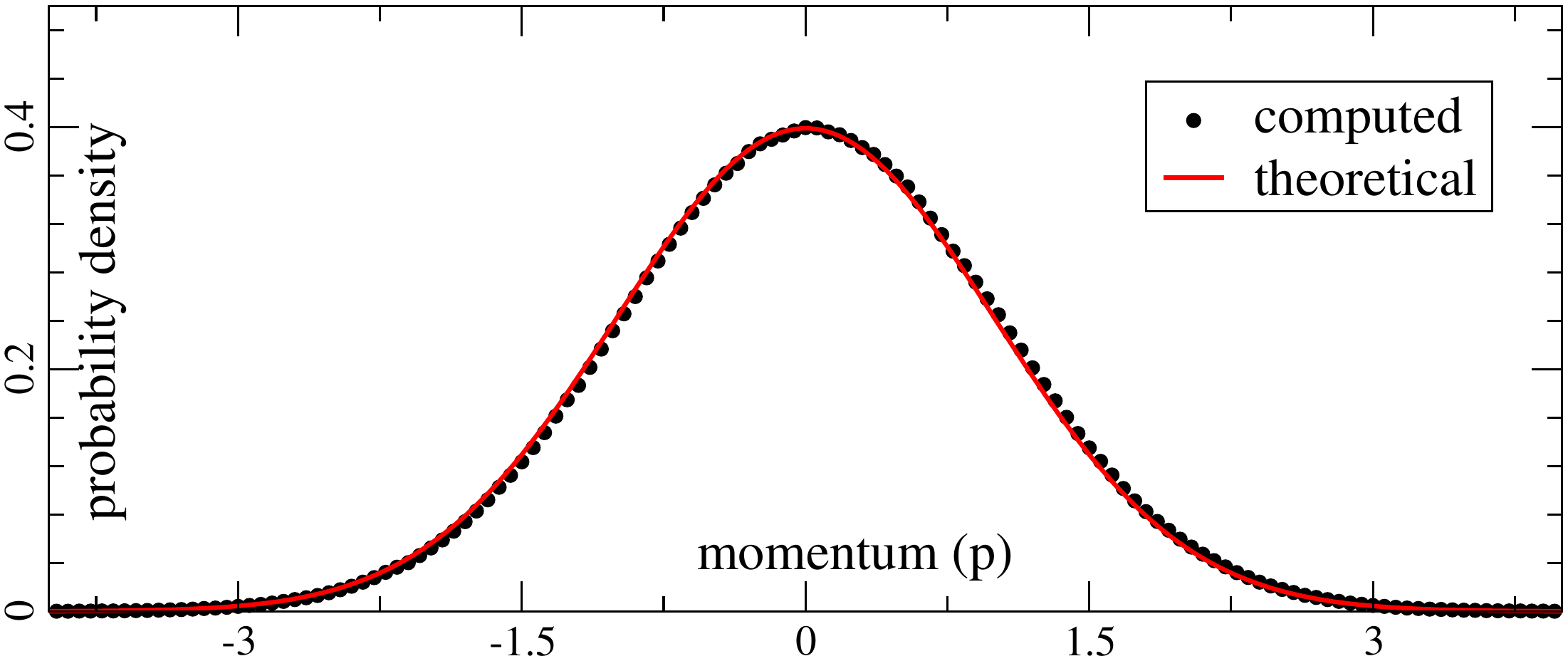}
\caption{\label{Fig.3} (Color online) The probability density of the momentum variable $p$ computed for the trajectory of length $t=10^{6}$ in numeric experiment 	{\bf{(f)}}=\{$\boldsymbol{\boldsymbol{\varphi}3},\mathbf{i2}$\}.   The theoretical curve is shown in red.}
\end{figure}

Fig.~\ref{Fig.1} and Fig.~\ref{Fig.2} clearly show that the joint temperature control for the systems $\mathrm{S}$ ans $\mathrm{S}^{*}$, which still involves a single extra variable, allows a practical ergodicity to be maintained. The trajectories {\bf{(b)}}, {\bf{(c)}}, {\bf{(e)}} and {\bf{(f)}} fill up the entire phase space and properly sample the canonical density. Dependence on initial conditions is not visually observed.
Fig.~\ref{Fig.3} shows coincidence of the computed and exact theoretical probability density curves for the momentum variable $p$ for the longer trajectory ($t=10^{6}$) generated in numerical experiment	{\bf{(f)}}=\{$\boldsymbol{\boldsymbol{\varphi}3},\mathbf{i2}$\} (selected solely for the sake of example).  For this trajectory, one can see that the computed probability density fits the exact theoretical curve perfectly. 

The results of numerical experiments presented in Figs.~\ref{Fig.1}
-\ref{Fig.3} give us the prospect for further detailing of the equations \eqref{eq:NH2} and \eqref{eq:NH2h} since the functions $\varphi_{p}(q)$ and $\varphi_{q}(p)$ are still of our choosing.
\section{Summary}
In this article, we have examined the importance of interpreting a deterministic thermostat as the part, $\mathrm{S^{*}}$, of the heat reservoir, $\Sigma $, that is involved in the joint dynamics with the physical system
under study, $\mathrm{S} $. The rest of the heat reservoir,  $\Sigma\setminus\mathrm{S}^{*}$, manifests itself only in the form of the parameter $\vartheta $, the temperature. 
In this framework, the thermal control of the system $\mathrm{S^{*}}$ should be considered on the same basis as the thermal control of the physical system $\mathrm{S} $. 
To this end, we have modified the NH scheme so that it uses a temperature control tool associated with the $\mathrm{S ^ {*}}$ system, which applies to all variables of the physical system $\mathrm{S} $ .
The modified equations of motion, which involve a single extra variable associated with the kinetic energy control, is derived as a special case of equations proposed in \citep{SamoletovVasiev2017}. With a single extra variable, we obtain the classical NH thermostat equations as a particular case of the modified equations, where all variables are subject to temperature control.

Then, we compare the classical NH equations and the modified equations in terms of their ergodicity in respect of the canonical distribution. We show that the modified equations manifest the ergodic property significantly better, while the computational costs involved remain minimal. For systems with many degrees of freedom, we expect the modified NH thermostat to be even more beneficial both in terms of ergodicity and computational costs. Indeed, in this case, the NH single extra variable controls only the total kinetic energy, but a complex molecular system can contain both fast and slow degrees of freedom or can be inhomogeneous such as to consist of weakly interacting subsystems, which can prevent the maintenance of the same temperature for all degrees of freedom. On the contrary, the modified NH dynamics provides simultaneous control of the temperature of all degrees of freedom with minimal computational costs. However, this remains a problem that needs further study.
\begin{acknowledgments}
This work has been supported by the EPSRC grant EP/N014499/1.
\end{acknowledgments}
\bibliography{thermo.bib}

\end{document}